\documentclass{article}
\usepackage[dvips]{graphics}
\begin{document}
\title{Computational Universality  and  $1/f$ Noise in Elementary Cellular Automata}
\author{Shigeru Ninagawa}
%

\maketitle

\begin{abstract}
It is speculated that there is a relationship between $1/f$ noise and computational universality in cellular automata. We use genetic algorithms to search for one-dimensional and two-state, five-neighbor cellular automata which have $1/f$-type spectrum. A power spectrum is calculated from the evolution starting from a random initial configuration. The fitness is estimated from the power spectrum in consideration of the similarity to $1/f$-type spectrum. The result shows that the rule with the highest average fitness has a propagating structure like other computationally universal cellular automata, although computational universality of the rule has not been proved yet.
\end{abstract}
\section{Introduction}
Cellular automata (CAs) is a  $d$-dimensional array with a finite automaton residing at each site. Each automaton called cell takes the states of neighbouring cells as input and makes the transition of its state according to a set of transition rules.
CAs are also considered to be spatially and temporally discrete dynamical systems with large degrees of freedom.
It was proved that elementary CA (ECA), namely one-dimensional and two-state, three-neighbor CA rule 110 is
computationally universal~\cite{Cook}. The ability to perform any algorithms is called computational universality.
In addition, the evolution of rule 110 starting from a random initial configuration
exhibits $1/f$ noise~\cite{Nina2}. Another example of computationally universal CA is the Game of Life (LIFE)~\cite{BCG}.
LIFE is a two-dimensional and two-state, nine-neighbor outer totalistic CA. It is supposed that LIFE is capable of supporting universal
computation, while the evolution starting from a random initial configuration has $1/f$-type spectrum~\cite{Nina}.
These results suggest that there is a relationship between computational universality and $1/f$ noise in CAs.

However the range of frequencies in the power spectra of ECA rule 110 that fits in power law is not broad.
This is caused by the periodic background that is a periodic pattern typically observed in the evolution of rule 110.
Since the periodic background does not play an essential role for performing computation,
we can guess that the essential feature of the evolution of rule 110 as a computing process is not lost by
the removal of the periodic background and that $1/f$-type spectrum becomes clear by removing  the periodic
background.

In this paper we study the influence of the removal  the periodic background on the shape of power spectrum of several
ECA rules.
In section 2 we explain the method of spectral analysis of elementary cellular automata.
In the following three sections we deals with ECA rule 110 as well as rule 54 and rule 62 which exhibit  power law
in power spectrum.
In the final section we discuss the relationship between computational universality and $1/f$ noise in CAs.

\section{Spectral analysis of cellular automata}

Let $s_x(t)  \in \{ 0, 1 \}$ be the value of site $x$ at time step $t$ in an elementary CA.
The site value evolves by iteration of the mapping,
\begin{equation}
\label{rule-func}
s_x(t+1) = F(s_{x-1}(t), s_x(t), s_{x+1}(t)).
\end{equation}
Here $F$ is an arbitrary function specifying the elementary CA rule. The elementary CA rule is determined by a binary
sequence with length $2^3=8$,
\begin{equation}
\label{rule-seq}
F(1,1,1), F(1,1,0),\cdots,F(0,0,0).
\end{equation}
Therefore the total number of possible distinct elementary CA rules is $2^8 = 256$ and each rule is abbreviated
by the decimal representation of the binary sequence as used in \cite{Wo1}.
Out of the 256 elementary CA rules 88 of them remain independent  (appendix of \cite{Li2}).

The discrete Fourier transformation of a time series of states $s_x(t)$ of site $x$ for $t = 0,1,\cdots,T-1$ is given by
\begin{equation}
\hat{s}_x(f) = \frac{1}{T}\sum_{t=0}^{T-1}s_x(t){\exp}
(-i\frac{2{\pi}tf}{T}) \hspace{10mm} f = 0,1,\cdots,T-1.
\end{equation}
The power per site is defined as
\begin{equation}
S(f) = \sum_{x=0}^{N-1} |\hat{s}_x(f)|^2,
\end{equation}
where $N$ denotes the total number of sites and the summation is taken in all sites.
The period of the component at a frequency $f$ in a power spectrum is given by $T/f$.
Throughout this paper we employ periodic boundary conditions where the sites are connected in a circle and
each array is started from a random initial configuration where each site takes state 0 or state 1 randomly with
independent equal probability.

The spectral analysis on the evolution of 88 independent ECAs starting from random initial configuration revealed that 
the power spectra of most of the rules is white noise or Lorenzian type and that rule 54 , rule 62 and rule 110 has the power
spectra of power law and especially rule 110 exhibits $1/f$ noise during the longest time steps~\cite{Nina2}.

we focus on the ECA rules which have power spectrum of power law, namely rule 54, rule 62 and rule 110.

\begin{figure}
\begin{center}
\scalebox{0.25}{\includegraphics{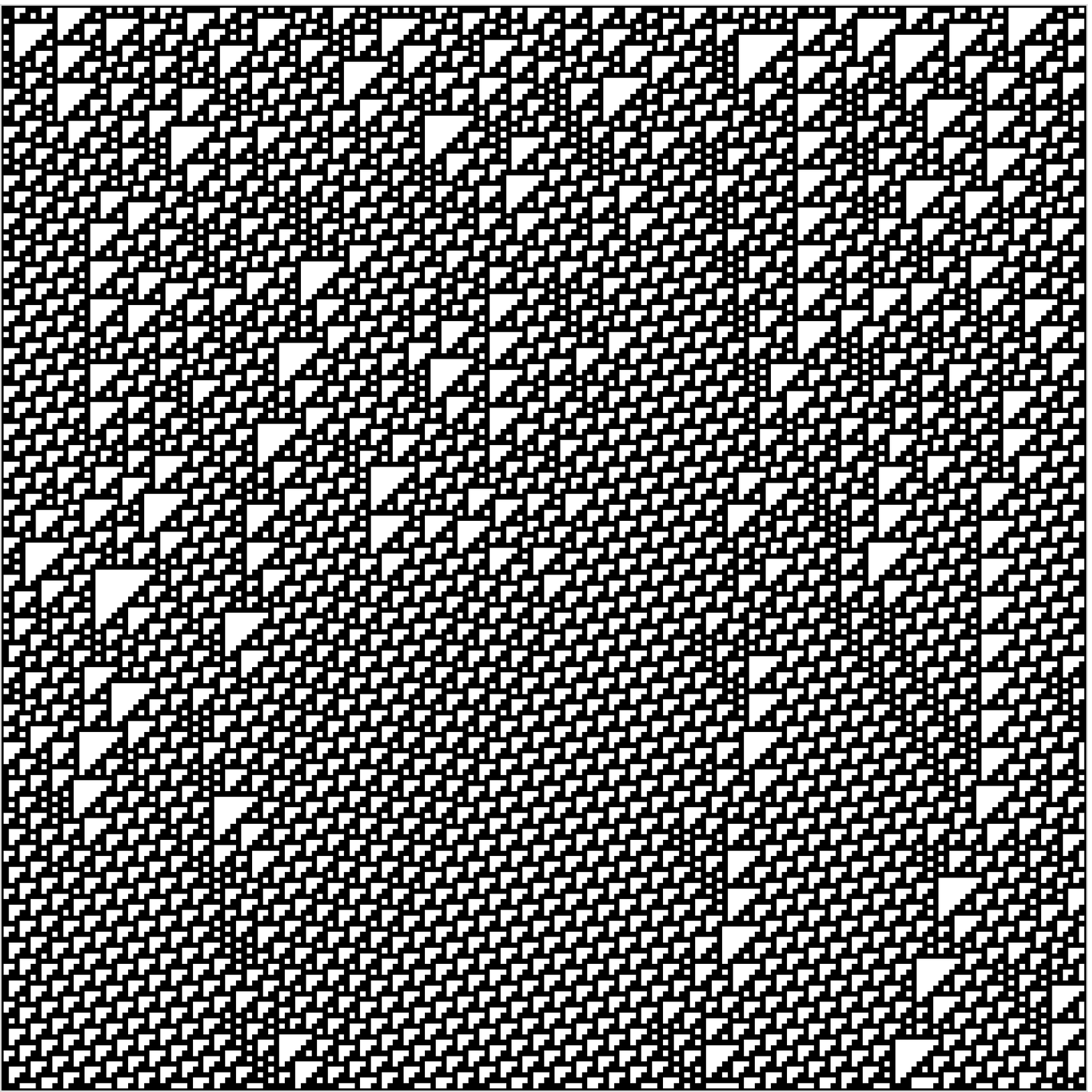}}
\scalebox{0.25}{\includegraphics{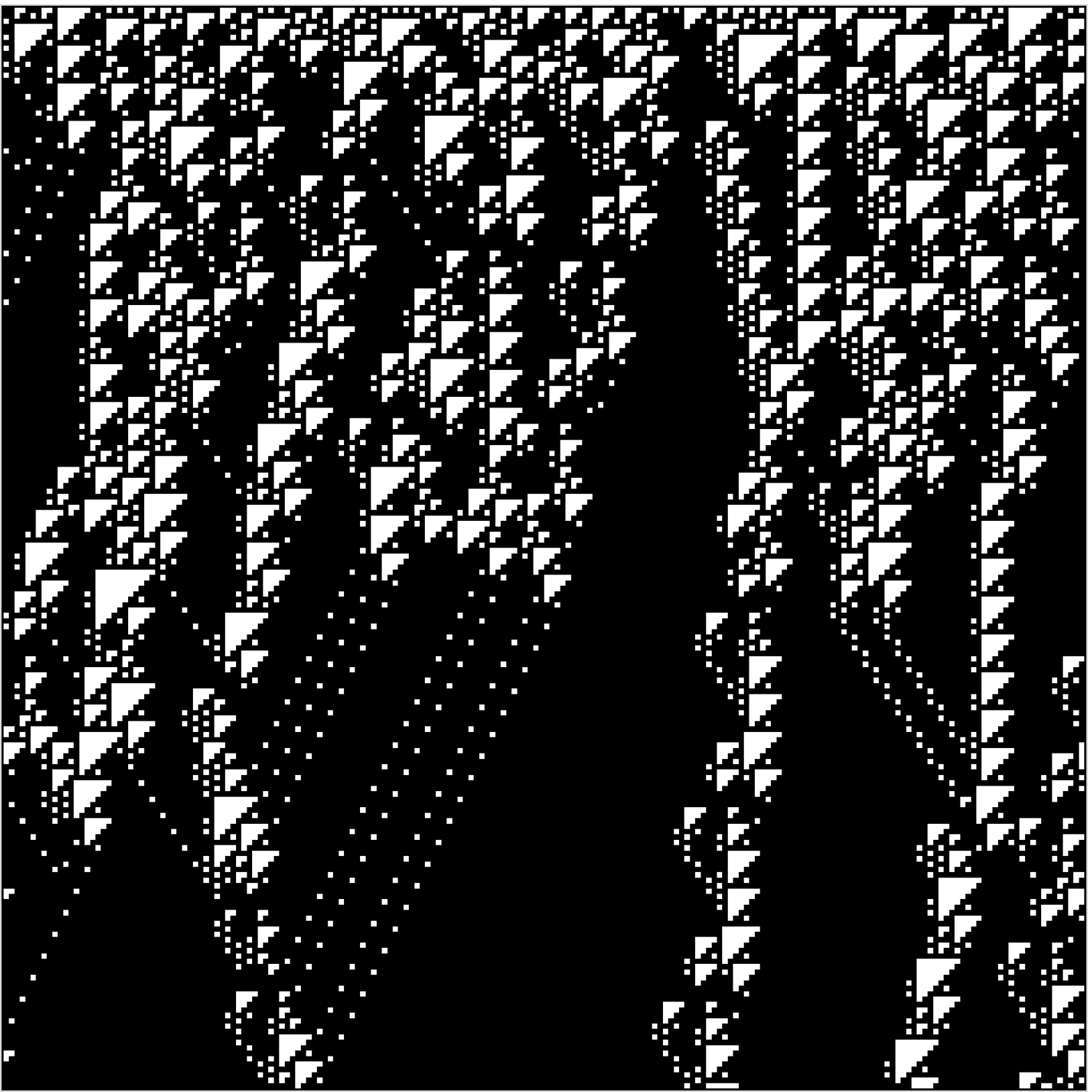}}
\end{center}
\caption{Space-time pattern (left) and the filtered one (right) of rule110 starting from a random initial configuration
with 200 cells for 200 time steps.}
\label{fig:SPTr110}
\end{figure}

\section{Rule 110}

The rule function of rule 110 is given by the following:
\[
\frac{111}{0} \frac{110}{1} \frac{101}{1} \frac{100}{0} \frac{011}{1} \frac{010}{1}
\frac{001}{1} \frac{000}{0}.
\]
The upper line represents the 8 possible states of neighborhood and the lower line specifies the state of the center cell at
the next time step.

\begin{figure}
\begin{center}
\scalebox{0.6}{\includegraphics{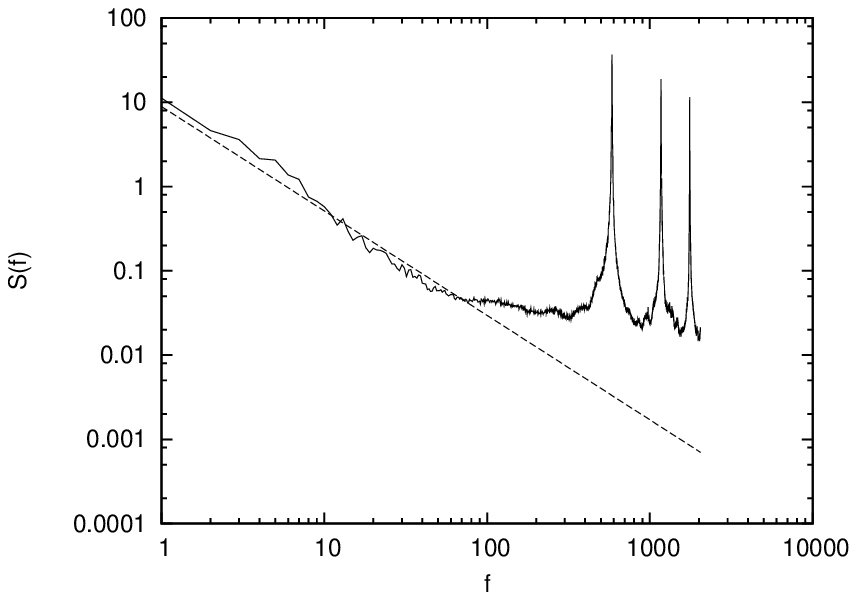}}
\scalebox{0.6}{\includegraphics{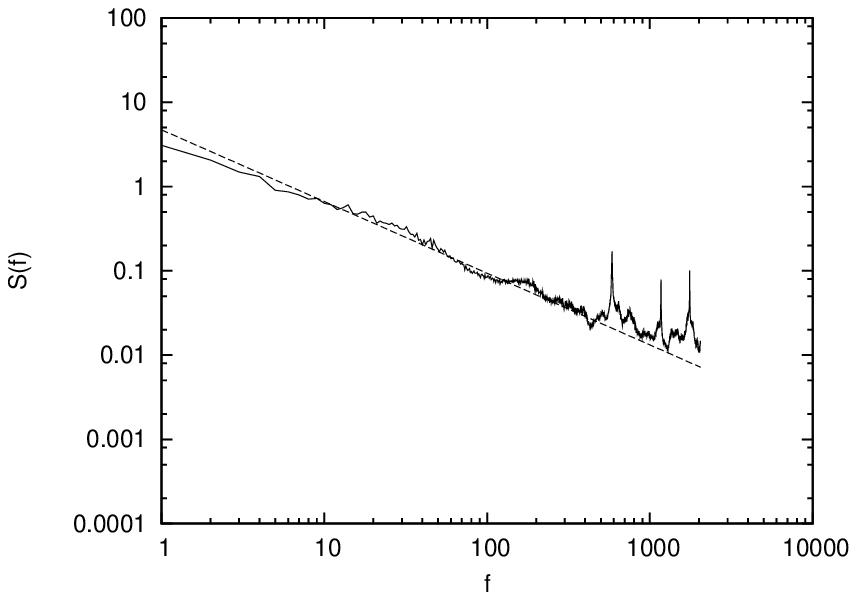}}
\end{center}
\caption{Left: Power spectrum of of rule 110.
The broken line represents the least-squares fitting of the power spectrum in the range of $f=1 \sim 100$ with the
exponent $\beta=-1.24$, the residual sum of squares $\sigma^2=0.047$.
Right: power spectrum calculated from the filtered space-time pattern ($\beta=-0.85$, $\sigma^2=0.020$).}
\label{fig:FSPr110}
\end{figure}

Figure \ref{fig:SPTr110} (left) shows a typical example of the space-time pattern of rule 110 starting from a random initial
configuration of 200 cells for 200 time steps. In space-time pattern, configurations obtained at successive time steps in the
evolution are shown on successive horizontal lines in which black squares represent sites with value 1, white squares sites
with value 0. We can observe periodic background of small white triangles with period seven in the space-time pattern.
Figure \ref{fig:FSPr110} (left) is the power spectrum of rule 110 calculated from the evolution starting from a random initial
configuration of 4000 cells for 4096 time steps. There are peaks at $f=585$ (period:7) and its harmonics in the spectrum.
The exponent $\beta$ of power spectrum is estimated by the least-squares fitting of the power spectrum $S(f)$
by $\ln(S(f)) = \alpha + \beta \ln(f)$.
The residual sum of squares $\sigma^2$ is given by
\begin{equation}
\sigma^2=\frac{1}{f_r}\sum_{f=1}^{f_r}{(\ln(S(f)) - \alpha - \beta \ln(f))^2},
\label{eq:residual}
\end{equation}
where $f_r$ is the number of data used for the calculation of $\sigma^2$ and we set $f_r=100$.
The broken line in the power spectrum represents the least-squares fitting of the power spectrum $S(f)$ in the range of $f=1 \sim 100$ with the exponent of power spectrum $\beta = -1.24$ and the residual
sum of squares $\sigma^2=0.047$.

\begin{figure}
\begin{center}
\scalebox{0.8}{\includegraphics{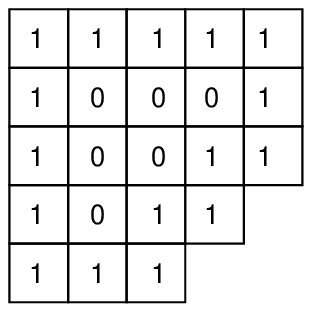}}
\end{center}
\caption{Template used to remove the periodic background of rule 110.}
\label{fig:pattern}
\end{figure}

The computational universality of rule 110 is proved by having rule 110 emulate cyclic tag system that is another
computationally universal system.
The stationary and propagating patterns and the interaction between those pattrens
are elaborately utilized for implementing cyclic tag system  in the evlution of rule 110.
On the other hand, the periodic background does not seem to play an essential role for performing computation.
Therefore we can guess that the essential feature of the evolution of rule 110 as a computing process is not lost by
the removal of the periodic background from the space-time pattern and that
the power spectrum calculated from the evolution obtained by removing  the periodic background
characterizes more vividly the behavior of rule 110 than the original power spectrum.

The periodic background has spatioal period 14 and temporal period 7 and 
consists of a template depicted in Fig. \ref{fig:pattern}.
To remove the periodic background from the space-time pattern, we scan the space-time pattern
from the upper left to the lower right searching for a section coincident with the template in Fig. \ref{fig:pattern} and
change the state zero into the state one in the matched section. By removing  the periodic background from
the original space-time pattern, we can get a filtered space-time pattern.

Figure \ref{fig:SPTr110} (right) shows the filtered space-tine pattern obtained 
from the original one in Fig. \ref{fig:SPTr110} (left) by means of the algorithms mentioned above.
The characteristic behavior in the evolution of rule 110 , that is,  the stationary and propagating patterns and
their interaction are readily apparent compared to the original space-time pattern.

Figure \ref{fig:FSPr110} (right) is the power spectrum calculated from the filtered space-time pattern obtained by 
removing the periodic background from the one which was employed to calculate the power spectrum shown in
Fig. \ref{fig:FSPr110} (left).
The peak at $f=585$ (period:7) and its harmonics considerably contract and
the spectrum seems to fit in power law in broader range of frequencies than the original one.
The exponent of power spectrum estimated by the least squares method in the range of frequencies $f=1 \sim 100$
is $\beta=-0.85$ and the residual sum of squares is $\sigma^2=0.020$.
These results imply that the removal of periodic background from the evolution of rule 110 brings about the effect that
the power spectrum comes close to power law.

\begin{figure}
\begin{center}
\scalebox{0.8}{\includegraphics{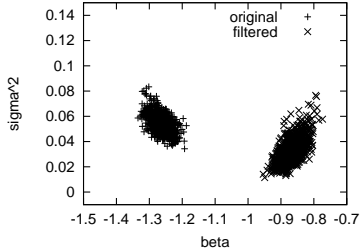}}
\end{center}
\caption{Scattergram of the exponents $\beta$ and the residual sum of squares $\sigma^2$ of 1000 power spectra
of rule 110 calculated from the original space-time patterns ($+$) and from the filtered ones ($\times$).}
\label{fig:fitR110}
\end{figure}

It is a matter of course that the power spectrum depends on initial configuration.
Next, let us investigate the statistical variation of the shape of power spectrum under the influence of the removal of
periodic backgroud in the evolution of rule 110.
We calculated the exponent $\beta$ and the residual sum of square $\sigma^2$ of
power spectrum in the range of frequencies $f=1 \sim 100 $ for 1000 runs starting from random initial configuration
of 4000 cells for 4096 time steps.
Figure \ref{fig:fitR110} is the scattergram of the exponents $\beta$ and the residual sum of squares $\sigma^2$ of
power spectra calculated from the original space-time patterns ($+$) and from the filtered ones ($\times$).
The pairs of parameters ( $\beta$, $\sigma^2$) of the original power spectra and the filtered ones
are evidently separated.
The $95\%$ confidence interval of population mean of $\beta$ and $\sigma^2$
are $\langle \beta \rangle = -1.264 \pm 0.001 $ and $\langle \sigma^2 \rangle = 0.0544 \pm 0.0004$
in original power spectra and $\langle \beta \rangle = -0.868 \pm 0.002$ and
$\langle \sigma^2 \rangle = 0.0324 \pm 0.0006$ in the filtered ones respectively.

\section{Other power law-type rules}
Rule 54 and rule 62 are two other rules which exhibit power law in power spectrum than rule 110 among 88 ECA rules.
In this section we study the property of these rules.

\begin{figure}
\begin{center}
\scalebox{0.2}{\includegraphics{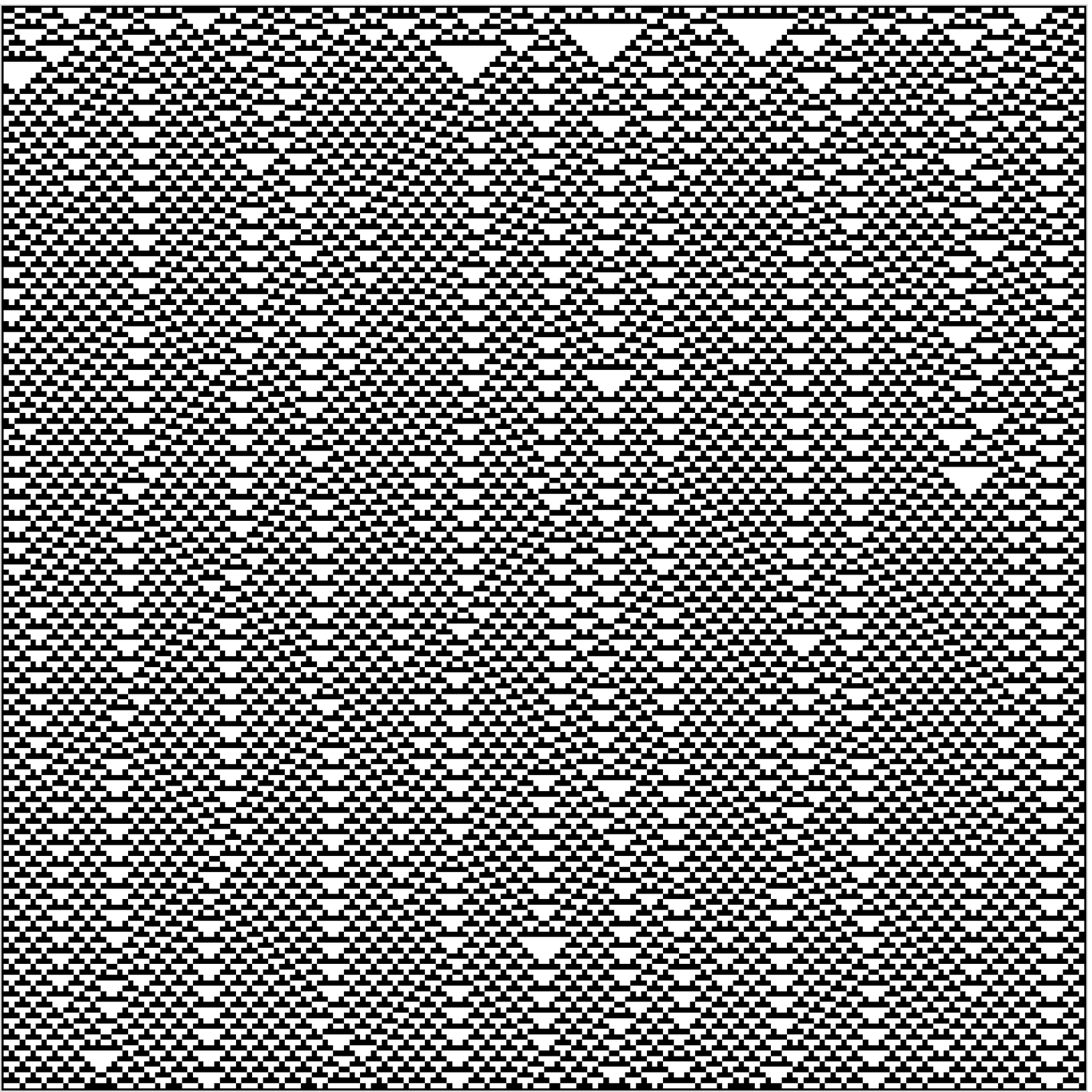}}
\scalebox{0.2}{\includegraphics{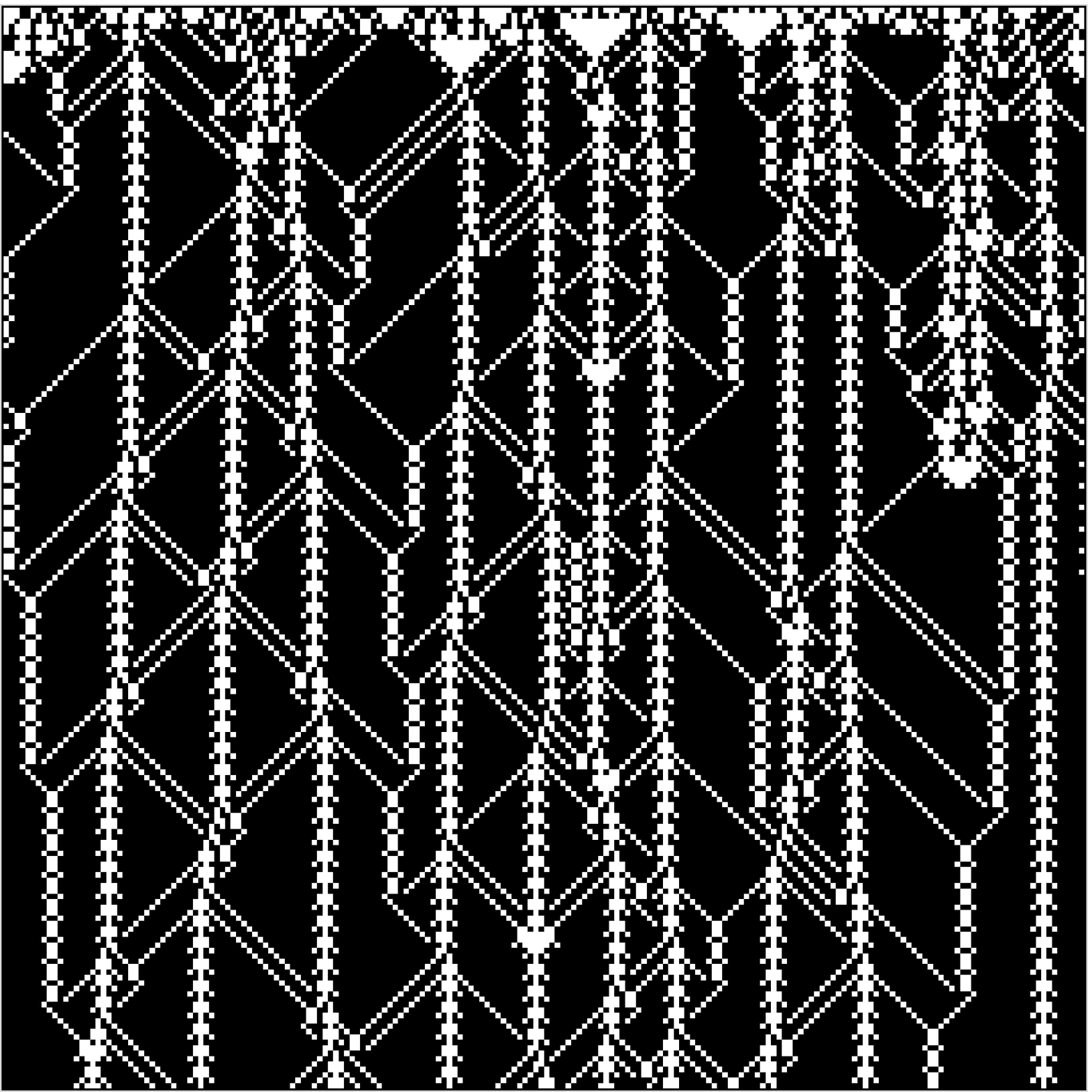}}
\end{center}
\caption{Space-time pattern (left) and the filtered one (right) of rule 54 starting from a random initial configuration
with 200 cells for 200 time steps.}
\label{fig:SPTr54}
\end{figure}

\begin{figure}
\begin{center}
\scalebox{0.55}{\includegraphics{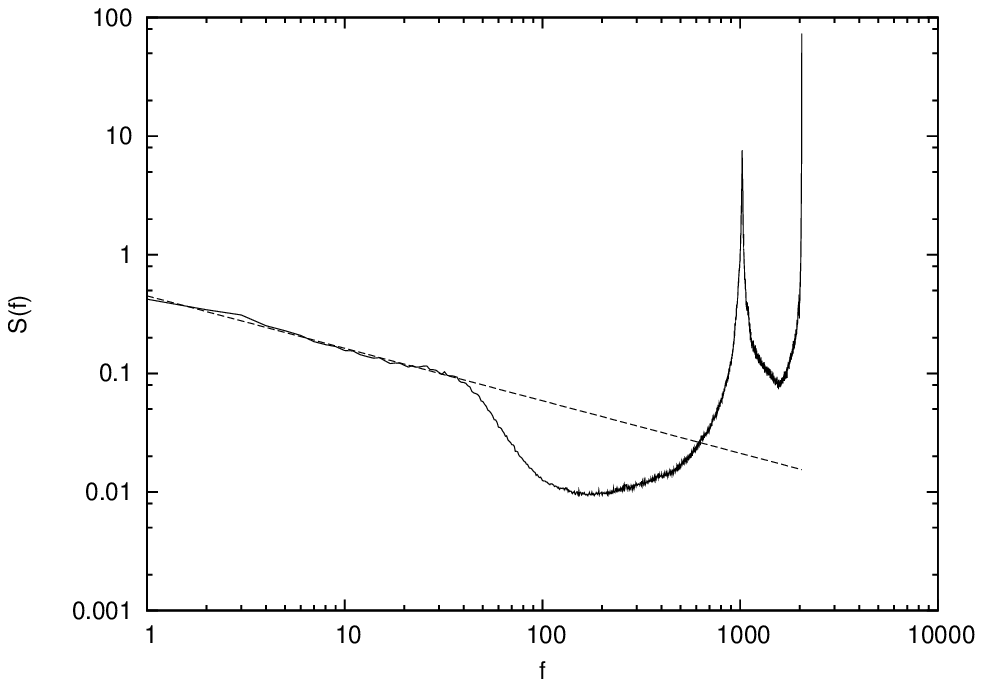}}
\scalebox{0.55}{\includegraphics{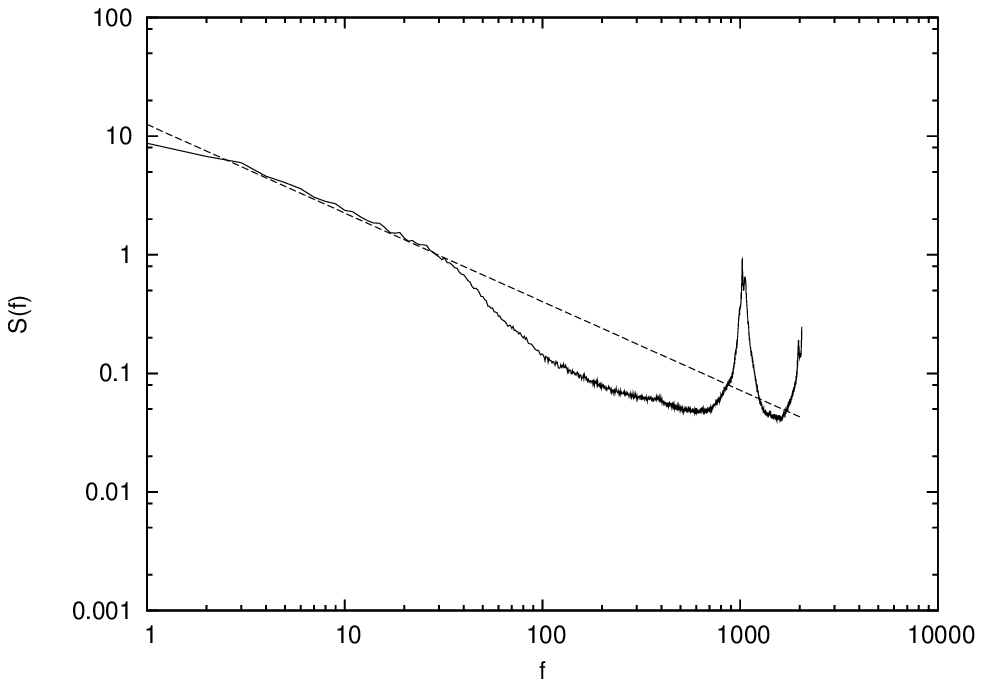}}
\end{center}
\caption{Left: Power spectrum of rule 54.
The broken line represents the least-squares fitting of the power spectrum in the
range of $f=1 \sim 100$ with the exponent $\beta= -0.909 $, the residual sum of squares $\sigma^2 = 0.149 $.
Right: power spectrum calculated from the filtered space-time pattern ($\beta= -1.086 $, $\sigma^2 = 0.063$).}
\label{fig:FSPr54}
\end{figure}

Figure \ref{fig:SPTr54} (left) shows the space-time pattern of rule 54 starting from a random initial configuration with
200 cells for 200 time steps. The periodic background of rule 54 has spatioal period 4 and temporal period 4.
Figure \ref{fig:FSPr54} (left) is the power spectrum of rule 54 calculated from the evolution starting from a random initial
configuration of 4000 cells for 4096 time steps. There are peaks at $f=1024$ (period:4) and its harmonics in the spectrum.
The exponent of the power spectrum estimated by the least squares method in the range of frequencies $f=1 \sim 100$
is $\beta = -0.909 $ and the residual sum of squares is $\sigma^2 = 0.149 $.

The filter to remove the periodic bakground of rule 54 is constructed by the following transfomation \cite{BNR}.
\begin{equation}
\sum_{i=0}^{3}s_{x+i}(t) \bmod 2 \rightarrow s_x(t).
\label{eq:BNRfilter}
\end{equation}

Figure \ref{fig:SPTr54} (right) shows the filtered space-tine pattern obtained 
from the left one. We can easily observe the stationary and propagating patterns and 
their interaction in the filterd space-time pattern in Fig. \ref{fig:SPTr54} (right).

Figure \ref{fig:FSPr54} (right) shows the power spectrum calculated from the filtered space-time pattern obtained by 
removing the periodic background from the one which was employed to calculate the power spectrum shown in
Fig. \ref{fig:FSPr54} (left).
The peak at $f=1024$ and its harmonics become blunt compared to the original one.
The exponent of power spectrum estimated by the least squares method in the range of frequencies $f=1 \sim 100$
is $\beta= -1.086 $ and the residual sum of squares is $\sigma^2 = 0.063 $.
These results imply that the removal of periodic background from the evolution of rule 110 brings about the effect that
the power spectrum comes close to 1/f-type.

\begin{figure}
\begin{center}
\scalebox{0.8}{\includegraphics{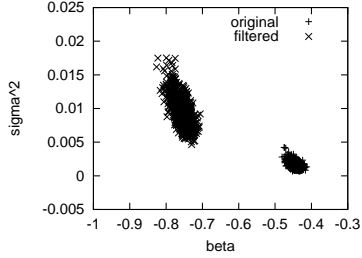}}
\end{center}
\caption{Scattergram of exponents $\beta$ and the residual sum of squares $\sigma^2$ of 1000 power spectra of rule 54
estimated from the original space-time patterns ($+$) and from the ones obtained by eliminating periodic
background ($\times$).}
\label{fig:fitR54}
\end{figure}

Figure \ref{fig:fitR54} is a scattergram of the exponents $\beta$ and the residual sum of squares $\sigma^2$ of
1000 power spectra of rule 54 calculated from the original space-time patterns ($+$) and from the filtered ones ($\times$).
The $95\%$ confidence interval of population mean of $\beta$ and $\sigma^2$
are $\langle \beta \rangle = -0.933 \pm 0.002 $ and $\langle \sigma^2 \rangle = 0.1551 \pm 0.0006 $
in original power spectra and $\langle \beta \rangle = -1.101 \pm 0.001 $ and
$\langle \sigma^2 \rangle = 0.0644 \pm 0.0001 $ in the filtered ones respectively.

\begin{figure}
\begin{center}
\scalebox{0.2}{\includegraphics{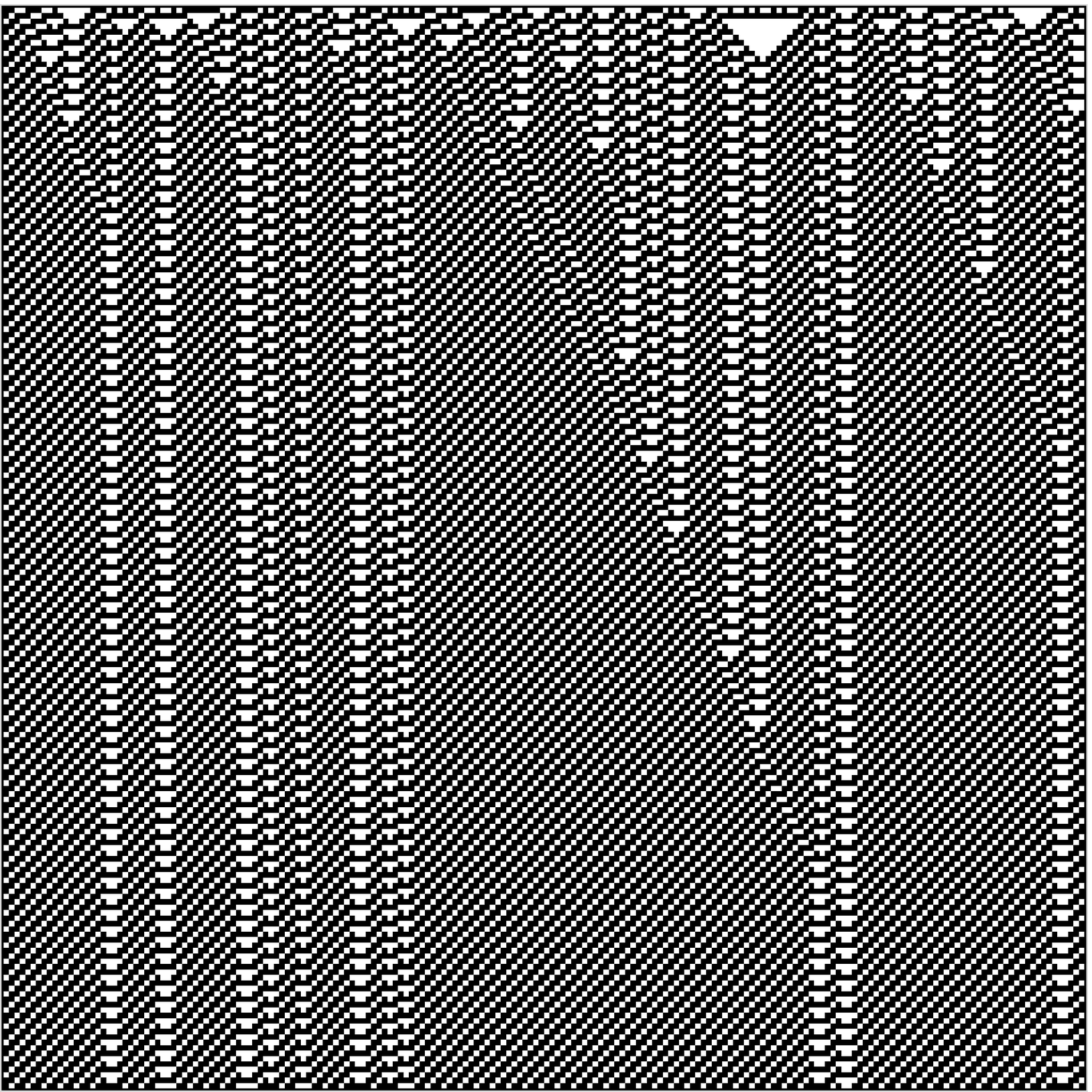}}
\scalebox{0.2}{\includegraphics{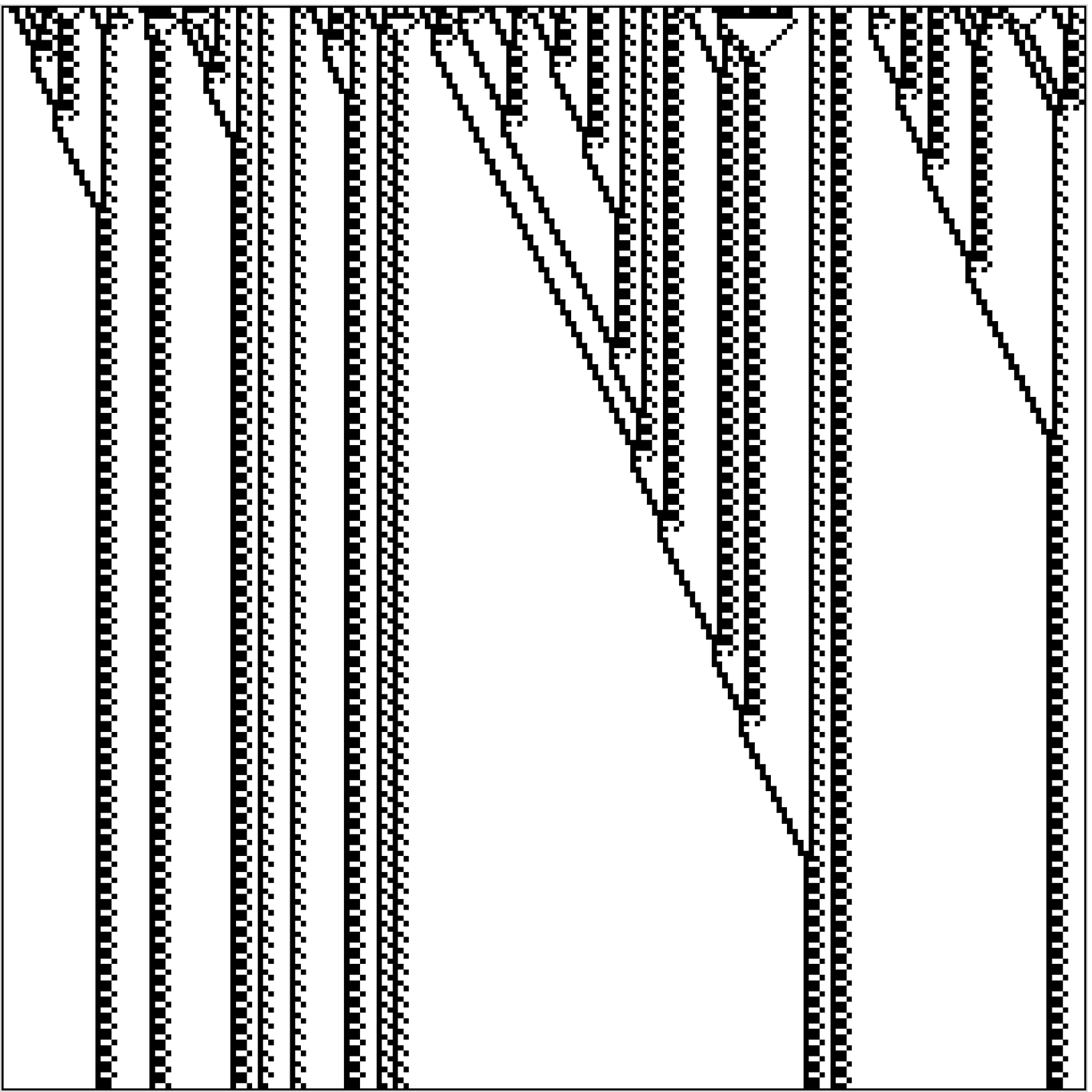}}
\end{center}
\caption{Space-time pattern (left) and the filtered one (right) of rule 62 starting from a random initial configuration
with 200 cells for 200 time steps.}
\label{fig:SPTr62}
\end{figure}

\begin{figure}
\begin{center}
\scalebox{0.55}{\includegraphics{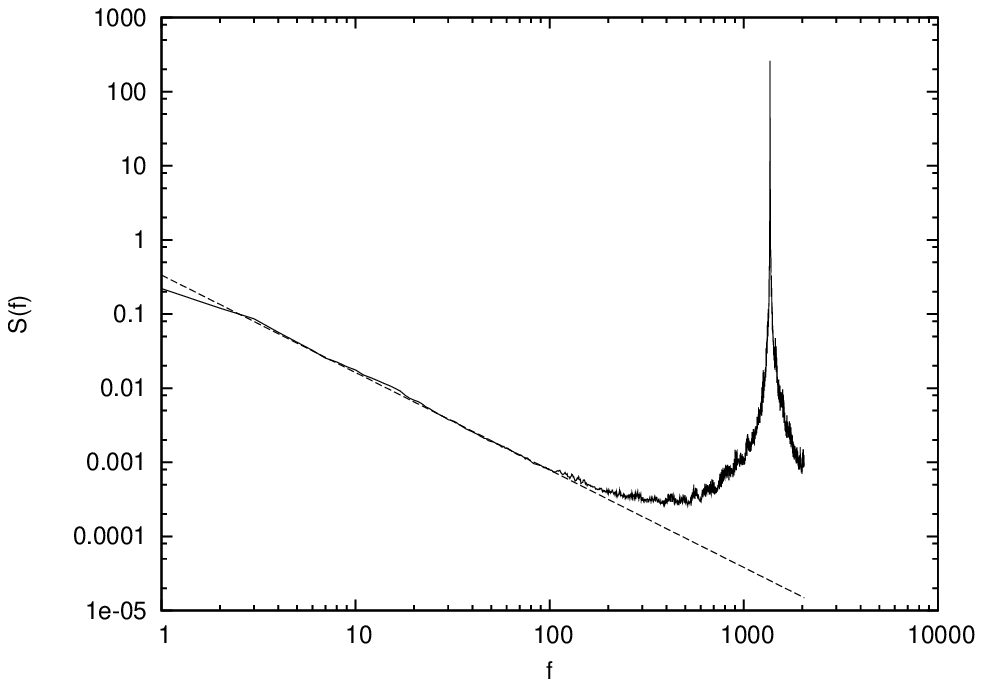}}
\scalebox{0.55}{\includegraphics{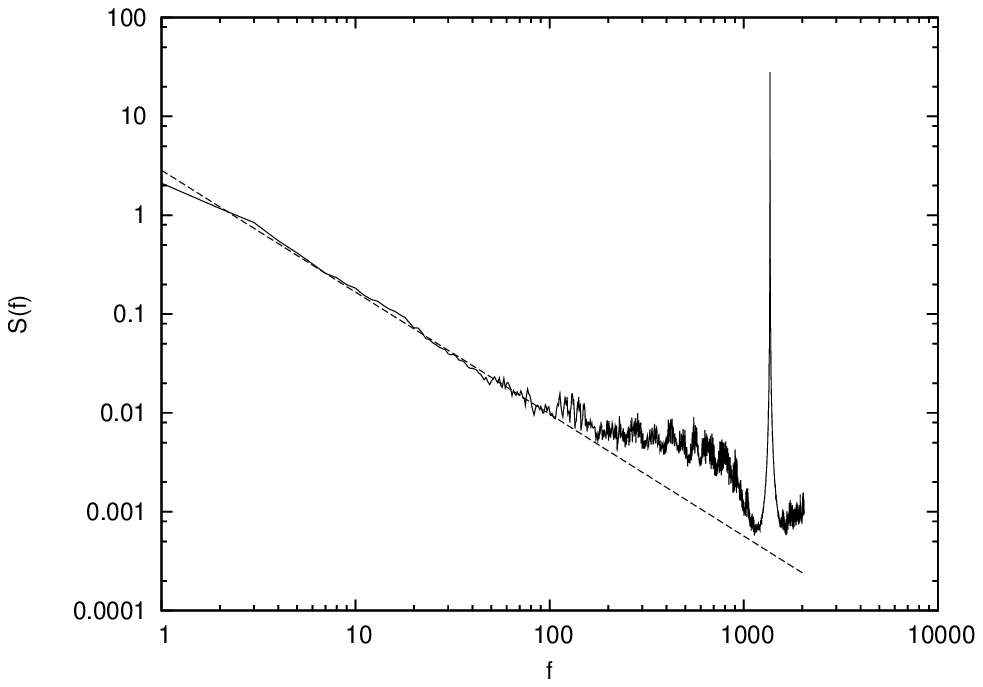}}
\end{center}
\caption{Power spectrum}
\caption{Left: Power spectrum of rule 62.
The broken line represents the least-squares fitting of the power spectrum in the
range of $f=1 \sim 100$ with the exponent $\beta= -1.315$, the residual sum of squares $\sigma^2 = 0.003 $.
Right: power spectrum calculated from the filtered space-time pattern ($\beta= -1.234 $, $\sigma^2 = 0.009$).}
\label{fig:FSPr62}
\end{figure}

Figure \ref{fig:SPTr62} (left) shows the space-time pattern of rule 62 starting from a random initial configuration with
200 cells for 200 time steps. The periodic background of rule 62 has spatioal period 3 and temporal period 3.
Figure \ref{fig:FSPr62} (left) is the power spectrum of rule 62 calculated from the evolution starting from a random initial
configuration of 4000 cells for 4096 time steps. There are peaks at $f=1365$ (period:3).
The exponent of the power spectrum estimated by the least squares method in the range of frequencies $f=1 \sim 100$
is $\beta = -1.315 $ and the residual sum of squares is $\sigma^2 = 0.003 $.

The filter to remove the periodic bakground of rule 62 is constructed by the following transfomation \cite{BNR}.
\begin{equation}
\sum_{i=0}^{2}s_{x+i}(t) \bmod 2 \rightarrow s_x(t).
\end{equation}

Figure \ref{fig:SPTr62} (right) shows the filtered space-tine pattern obtained 
from the left one. We can easily observe the stationary and propagating patterns and 
their interaction in the filterd space-time pattern in Fig. \ref{fig:SPTr62} (right).

Figure \ref{fig:FSPr62} (right) shows the power spectrum calculated from the filtered space-time pattern obtained by 
removing the periodic background from the one which was employed to calculate the power spectrum shown in
Fig. \ref{fig:FSPr62} (left).
The exponent of power spectrum estimated by the least squares method in the range of frequencies $f=1 \sim 100$
is $\beta = -1.234$ and the residual sum of squares is $\sigma^2 = 0.009$.

\begin{figure}
\begin{center}
\scalebox{0.6}{\includegraphics{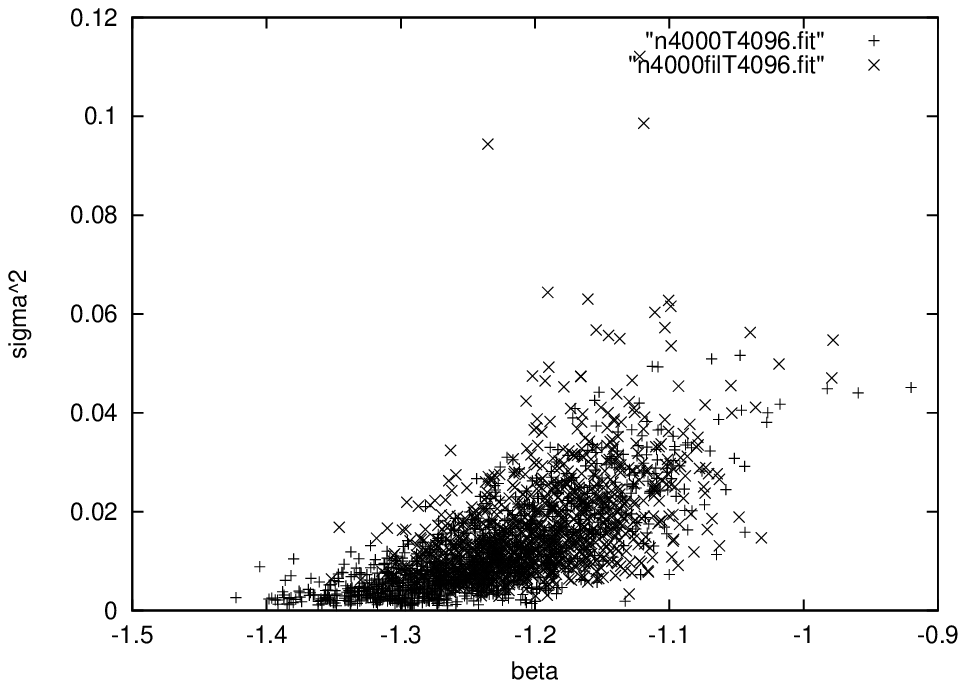}}
\scalebox{0.6}{\includegraphics{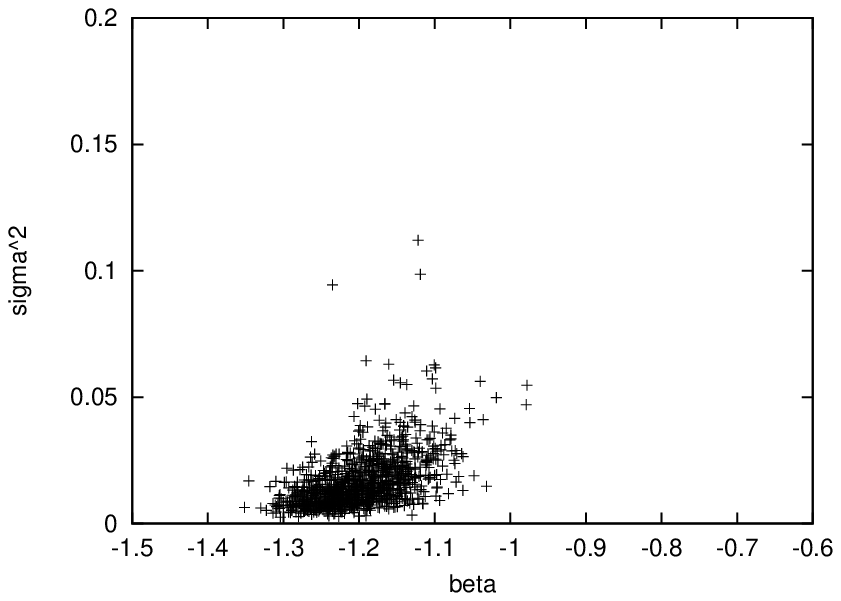}}
\end{center}
\caption{Scattergram of exponents $\beta$ and the residual sum of squares $\sigma^2$ of 1000 power spectra of rule 62
estimated from the original space-time patterns (left) and from the filtered ones (right).}
\label{fig:fitR62}
\end{figure}

Figure \ref{fig:fitR62} is a scattergram of the exponents $\beta$ and the residual sum of squares $\sigma^2$ of
1000 power spectra of rule 62 calculated from the original space-time patterns ($+$) and from the filtered ones ($\times$).
The $95\%$ confidence interval of population mean of $\beta$ and $\sigma^2$
are $\langle \beta \rangle = -0.933 \pm 0.002 $ and $\langle \sigma^2 \rangle = 0.1551 \pm 0.0006 $
in original power spectra and $\langle \beta \rangle = -1.101 \pm 0.001 $ and
$\langle \sigma^2 \rangle = 0.0644 \pm 0.0001 $ in the filtered ones respectively.
It is difficult to separate the data into two groups.

\section{Discussion}
Generally speaking, the process capable of supporting computation needs the three kinds of functions on information, that is,
transmission, storage, and operation of information.
The transmission of information is achieved by propagating patterns and the storage is done by stationary patterns
while the operation of information is carried on by the interaction between those patterns.

It is believed that rule 62 in not capable of supporting universal computation because the way in which information is
transmitted is highly constrained~\cite{NKS}.

As an additional experiment, we calculated the exponent $\beta$ of power spectrum under the same conditions except for
the time steps, $T=1024$ and 2048.
The $95\%$ confidence interval of population mean of $\beta$ are $\langle \beta \rangle = -0.6196 \pm 0.0005 $ for $T=1024$
$\langle \beta \rangle = -0.9229 \pm 0.0008 $ for $T=2048$ in original power spectra and
$\langle \beta \rangle = -0.8203 \pm 0.0007 $ for $T=1024$ $\langle \beta \rangle = -0.821 \pm 0.001 $ for $T=2048$
in filtered ones.
In the original power spectra $|\beta|$ becomes large as the time steps $T$ gets long while $\beta$ hardly varies with $T$
in the filtered power spectra.

%
%

\end{document}